# Closing the Gap between Teaching and Assessment

Chandralekha Singh

Department of Physics and Astronomy and Discipline-based Science Education Research Center (dB-SERC)

University of Pittsburgh, Pittsburgh, PA 15260

Evidence-based teaching is based upon a model of learning in which assessment plays a central role [1]. New knowledge necessarily builds on prior knowledge [2-3]. According to the model shown schematically in the figure below, students enroll in a course with some initial knowledge relevant for the course. The instruction should be designed carefully to build on the initial knowledge and take students from that initial knowledge state to a final knowledge state based upon the course goals [3]. It is difficult to know what students know at the beginning or at the end of a course. Assessment is the process in which both students and instructor get feedback on what students have learned and what is their level of understanding vis a vis the course goals [1-5]. To assess learning, students can be given pre-tests and post-tests at the beginning and at the end of the course [6-17]. The performance on these tests can reflect the extent to which the course goals have been achieved if the tests are designed carefully consistent with the course goals to reflect students' knowledge state accurately [6-17].

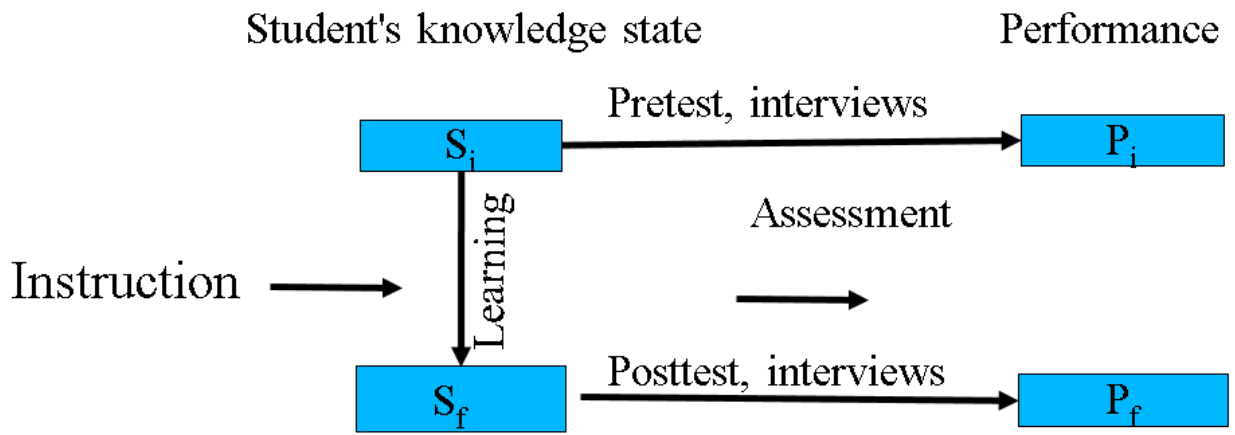

Research suggests that students must construct their own understanding though the instructor plays a critical role in helping students accomplish this task [1-3, 18-19]. An instructor should model the criteria of good performance, while leaving sufficient time to provide guidance and feedback to students as they practice useful skills. The amount of instructional support given to students should be decreased gradually as they develop self-reliance. A particularly effective approach is to let students work in small groups and take advantage of each other's strengths [20-23].

Discipline-based education research emphasizes the importance of having well-defined learning goals and assessing student learning using tools that are commensurate with those goals. "Students should understand acceleration" is not a well-defined goal because it does not make it clear to students what they should be able to do. Students may misinterpret this instructor's goal to imply that they should know the definition of acceleration as the rate of change of velocity with time, while the instructor may expect students to be able to calculate the acceleration when the initial and final velocities of an object and the elapsed time are provided. Examples of measurable goals should be shared with students and should include cognitive achievement at all levels of Bloom's taxonomy including students demonstrating how to apply concepts in diverse situations, analyzing problems by breaking them down into sub-problems, synthesizing a solution by combining different principles, and comparing and contrasting various concepts.

Assessment drives learning and students focus on learning what they are tested on. Therefore, using assessment tools that only probe mastery of algorithms and plug and chug approaches will eliminate incentives to acquire deep understanding [1-3, 24]. Furthermore, instructional design should be targeted at a level where students struggle appropriately and stay engaged in the learning process and the material should not be so unfamiliar and advanced that students become frustrated and disengage [25-33]. Research-based curricula and pedagogies are designed to take into account the initial knowledge of a typical student and gradually build on it.

The course assessments should be viewed as formative, i.e., conducive to student learning, rather than summative. If used frequently throughout the course, formative assessments can greatly improve student learning since they have many opportunities to reflect on their learning consistent with course goals. In addition, frequent assessment can help the instructor get real time feedback on the effectiveness of instruction which can be used to refine instruction and address difficulties [1].

Therefore, formative assessment, e.g., think-pair-share activities, clicker questions, tutorials with pre- and post-tests, collaborative problem solving, process oriented guided inquiry learning, minute papers, asking students to summarize what they learned in each class or asking them to make concept maps of related concepts etc. should be used throughout the course in order to help students build a good knowledge structure and develop useful skills [1-3, 18]. Using these low-stakes formative assessment tools that are "built-in" or integrated with the rest of the instructional material throughout the course can bridge the gap between teaching and assessment.

To summarize, the formative assessment approaches used throughout the course at least in part accomplish the following:

- They provide students with an understanding of the goals of the course in a concrete manner with specific examples since the activities that they engage in communicate instructor's expectations (I expect that you are able to solve this type of problems, complete these types of tasks etc.)
- They provide students with feedback on where their understanding is at a given time with relation to the course goals as communicated by the instructor
- the timely feedback students obtain helps them improve their understanding early when there is time to catch up
- They provide the instructor with feedback on where the class is with relation to the course goals
- They keep students active in the learning process throughout the course and provide instructors an opportunity to coach students appropriately and take advantage of each other's strengths

The above discussions suggest an important aspect of formative assessment: significant use of formative assessment tools which are carefully embedded and integrated in the entire instruction entails a close scrutiny of all aspects of an instructional design. Before implementing evidence-based teaching and learning one should compile a list of what initial knowledge students have, what are the measurable goals of the course and think carefully about how to design instruction aligned with the initial knowledge of students and course goals and how to scientifically assess the extent to which each measurable goal is achieved. In essence, evidence-based teaching entails that instructors carefully contemplate the answers to the following questions:

1. What initial knowledge and cognitive and affective attributes do students have that is relevant for instruction (content-based initial knowledge including alternative-conceptions, problem solving and reasoning skills, mathematical skills, epistemological beliefs, attitude, motivation, self-efficacy etc.)?
2. What should students know and be able to do?
3. What does proficiency in various components of this course look like consistent with goals?
4. What evidence would I accept as demonstrating proficiency? What evidence would be acceptable to most of my colleagues?
5. How can I design instruction that builds on students' initial knowledge and takes them systematically to a final knowledge state which is commensurate with course goals?
6. Are the initial knowledge of students, course goals, instructional design and assessment methods aligned with each other?

A typical goal of a science course is to provide students with a firm conceptual understanding of the underlying knowledge. Therefore, discipline-based education researchers in many disciplines have developed assessment instruments designed to assess students' conceptual understanding. The data from these instruments can be used to assess the initial knowledge of the students if administered as a pre-test before instruction and inform what students learned and what aspects of the material were challenging if given after instruction as a post-test. These data can help improve instructional design, e.g., by pinpointing where more attention should be focused to improve student learning. If the post-test is administered right after instruction in a particular course, it can serve as a formative assessment tool and if it is given at the end of the term, it can be helpful for improving learning for future students.

Another course goal may be to improve students' attitudes about the nature of science and learning science and provide them with an understanding of what science *is* and what it takes to be successful in science courses [34]. To this end as well, discipline-based educational researchers have developed assessment instruments that are either discipline specific or about science in general. Also, in many science courses, students are expected to develop good problem solving strategies and to this end, instruments have been developed to assess students' attitudes and approaches to problem solving.

The standardized assessment instruments that have been developed for science courses provide a starting point for thinking about assessing effectiveness of teaching and learning and for investigating the extent to which various instructional goals have been met [e.g., see Ref. [7]-[17] for examples of standardized assessment tools for introductory physics]. The data obtained from these instruments can be compared with national norms found in publications in education research journals. See http://www.dbserc.pitt.edu for examples of such instruments in natural sciences, e.g., physics, chemistry, biology, mathematics etc.